\documentclass[fleqn,usenatbib]{mnras}

\usepackage[T1]{fontenc}
\usepackage{ae,aecompl}

\usepackage{graphicx}	
\usepackage{amsmath}	
\usepackage{amssymb}	
\usepackage{courier}
\usepackage{xcolor}

\newcommand\be{\begin{equation}}
\newcommand\ee{\end{equation}}

\newcommand\ikx{i{\bf k}_p\cdot {\bf x}_p}
\newcommand\tpz{\tilde{\Phi}_1}
\newcommand\kx{{\bf x}_p\cdot {\bf k}_p}

\newcommand\kms{{\rm km\,s}^{-1}}
\newcommand\kmskpc{{\rm km\,s}^{-1}\,{\rm kpc}^{-1}}
\usepackage{hyperref}
\hypersetup{colorlinks=true,linkcolor=blue,filecolor=magenta,urlcolor=blue}

\title[Swing-Amplified Phase Spirals]{Swing Amplification and the Gaia Phase Spirals}

\author[Lawrence M. Widrow]{Lawrence M. Widrow\thanks{E-mail: widrow@queensu.ca}
\\
Department of Physics, Engineering Physics and Astronomy, Queen's University, Kingston, K7L 3X5, Canada}

\date{Accepted XXX. Received YYY; in original form ZZZ}

\pubyear{2020}

\begin{document}
\label{firstpage}
\pagerange{\pageref{firstpage}--\pageref{lastpage}}
\maketitle

\begin{abstract}
We explore the interplay between in-plane and vertical dynamics in stellar discs within the framework of the shearing box approximation. Julian and Toomre used the shearing sheet to show that leading density waves are amplified as they swing into a trailing ones. We extend their formalism into the dimension perpendicular to the disc and obtain explicit solutions for the response of a disc to an impulsive, external excitation. An excitation that is is symmetric about the mid plane produces a density/breathing wave as well as two-armed phase spirals in the vertical phase space plane. On the other hand, an excitation that is antisymmetric about the mid plane leads to a bending wave and single-armed phase spirals. In either case, self-gravity plays a crucial role in driving the evolution of the disturbance and determining the amplitude and pitch angle of the ensuing spirals. We also show that when the disc is excited by a co-rotating cloud, it develops stationary phase spirals in the wake of the cloud. The results call into question simple kinematic arguments that have been used to determine the age of the phase spirals seen in the Gaia survey.

\end{abstract}



\section{Introduction}\label{intro}

One of the most intriguing discoveries from Gaia Data Release 2 \citep{gaiadr2_summary, gaiadr2_vrad} is the existence of spirals in the vertical, or $z-w$, phase space distribution function (DF) of Solar Neighbourhood stars \citep{antoja2018}. The phase spirals are easily seen in maps of the $z-w$ DF once the smooth background distribution has been removed. They have a fractional density contrast of a few percent and display a rich morphology that depends on the properties of the stars under consideration such as Galactocentric radius, Galactic azimuth, angular momentum, and epicyclic energy \citep{laporte2019, Widmark2019, blandhawthorn2019, Li2020, hunt2022, frankel2022, antoja2022}. For example the spirals tend to be two-armed in the inner galaxy and one-armed in the outer galaxy \citep{hunt2022}. They also appear in $z-w$ maps of the mean azimuthal and radial velocities.

The most natural explanation for the spirals is that they are disturbances in the $z-w$ DF from some past event or events that have undergone phase mixing \citep{tremaine1999}. For example, if a local patch of the disc experiences a "kick" perpendicular to the mid plane, stars will be displaced in the $w$-direction. The perturbed DF will then shear into a one-armed spiral since stars with low vertical energy rotate about the origin of the $z-w$ plane at a higher frequency than stars with high vertical energy. On the other hand, one might imagine a breathing mode perturbation where the DF is squeezed in $z$ and/or stretched in $w$. Over time, this perturbation will shear into a two-armed spiral. In either case, if the evolution in phase space is purely kinematic, then the pitch angle of the phase spiral will depend on the time since the initial perturbation and the anharmonicity of the vertical potential. Indeed, the simplest approach to understanding the Gaia phase spirals is to model the DF as test particles in a fixed one-dimensional potential, introduce an ad hoc perturbation, and evolve the system until a spiral pattern matching the one seen in the data is reached. This approach leads to an estimate of $300-900\,{\rm Myr}$ for the age of the spiral \citep{antoja2018}. When the test-particle analysis is extended to three-dimensions, it can help elucidate the origin of the mean $v_\phi$ and $v_R$ spirals \citep{binney2018, darling2019a}.

Similar estimates for the age of the spirals can be obtained by transforming the data into action-angle-frequency or $(J_z,\,\theta_z,\,\Omega_z)$ coordinates. If the spirals are created by a single event and if the potential is time-independent, then $J_z$ will be constant and $\theta_z = \Omega_z t + \theta_0$ along a star's orbit. Thus, in the $\theta_z-\Omega_z$ plane, the spirals should appear as parallel ridges whose slope is proportional to the inverse age of the spiral \citep{LW2021, frankel2022, LW2023, tremaine2022}. Note that both methods require a model for the background gravitational potential. An alternative approach is to use the shape of the spirals to constrain the gravitational potential \citep{widmark2021a, widmark2021b}.

A promising candidate for the origin of the spirals is the passage of a dwarf galaxy or dark matter subhalo through the Galactic disc with the Sagittarius dwarf galaxy (Sgr) considered a prime suspect \citep{laporte2019, bennett2021, bennett2022}. Sgr is the nearest known neighbor to the Milky Way \citep{ibata1994} and has an orbit that has likely taken it through the mid plane of the Galaxy several times over the last few Gyr \citep{johnston1995}. \citet{purcell2011} argued that Sgr was crucial in shaping the Milky Way's bar and spiral arms while \citet{gomez2012} suggested that it could have also generated the vertical bending and breathing waves seen in both pre-Gaia surveys and Gaia \citep{widrow2012, williams2013, carlin2013, xu2015, bennett2019}. Several groups have found vertical phase spirals in high-resolution simulations of a Sgr-Milky Way encounter though none of these have managed to reproduce the morphology of the spirals found in the Gaia data \citep{laporte2019, bennett2021}. Simulations have also been used to explore other origins of the phase spirals such as the vertical waves generated by a buckling event in the Galactic bar \citep{khoperskov2019}. It is worth noting the spirals are subtle (few percent) phase space features at a scale of $200\,{\rm pc}$ by $10\,\kms$ and therefore push the limits of the resolution simulations.

The general conclusion from these investigations is that the simple picture of a kinematic spiral generated from a single event is incomplete, if not incorrect. For example, the transition from two-armed to one-armed spirals as one moves out in Galactocentric radius may require multiple events \citep{hunt2022}. In addition, when the spirals are transformed from $z-w$ to $\theta_z-\Omega_z$ coordinates, they appear as curved rather than parallel bands \citep{frankel2022,tremaine2022}. Finally the phase spirals found in simulations are not as tightly wound as the ones seen in the data \citep{laporte2019, bennett2021, bennett2022}. These results may indicate that self-gravity is essential for modelling the evolution of the spirals. This point was stressed in \citet{darling2019a} who compared N-body simulations of a test-particle disc with fully self-consistent ones. In both cases, a bend at the solar circle was introduced into the disc. The spirals that developed in the test-particle case were easy to detect and had the expected pitch angle. On the other hand, the spirals in the live disc were less tightly wound and more difficult to discern.

Recently, \citet{tremaine2022} proposed an alternative scenario in which phase spirals are generated by a continual sequence of weak perturbations and erased by phase space diffusion due to the graininess of the gravitational potential. In this picture, the pitch angle of the spirals reflects the diffusion time scale rather than the elapsed time from an initial perturbation. As a proof of concept, they presented simulations in which test particles in a fixed potential were subjected to a stochastic sequence of kicks and were able to reproduce key features of the Gaia spirals.

In this paper we explore the connection between in-plane perturbations and phase spirals within the framework of the shearing box. In their classic paper on swing amplification, \citet{julian1966} (hereafter, JT66) followed the evolution of a plane wave perturbation in a razor-thin disc by integrating the linearized equations for the surface density and gravitational potential. They found that in a marginally stable disc (Toomre parameter $Q\ga 1$) with surface density $\Sigma_0$ and epicyclic frequency $\kappa$, waves with wavelength close to $\lambda_{\rm crit} = 4\pi^2 G\Sigma_0/\kappa^2$ were amplified by one or two orders of magnitude as they swung from leading to trailing. Our first task will be to extend the JT66 formalism into the dimension perpendicular to the disc, that is, to go from a shearing sheet to a shearing box. The shearing box approximation has been used in a variety of problems in theoretical astrophysics such as the study of accretion discs and, most notably, the magnetorotational instability \citep{hawley1995}. There, the evolution of the system is driven by gasdynamics and magnetic fields. In our case, the system is collisionless and the evolution is driven entirely by gravity.

As in JT66, we solve the linearized collisionless Boltzmann equation for a disc that is perturbed by an external excitation. The approach has some commonalities with the formalism developed in \citet{banik2022} who also considered the response of an isothermal slab to an external potential in linear theory. However, their analysis did not include self-gravity, epicyclic motions, or shear, all of which play important roles in the our analysis. We also build on early studies of self-gravitating modes in plane-symmetric systems by \citet{kalnajs1973, araki1985, mathur1990, weinberg1991, widrow2015}.

Though this work focuses on the $z-w$ phase spirals, the shearing box machinery can be used to investigate more general questions about the interplay between in-plane and vertical dynamics. In particular, one can study the relationship between in-plane disturbances such as spiral arms and the vertical waves seen throughout the Milky Way's disc \citep{widrow2012,carlin2013,williams2013, yanny2013, xu2015, schonrich2018, bennett2019,widmark2022}. The connection between vertical breathing waves and spiral arms was investigated with N-body simulations by \citet{debattista2014,ghosh2022,kumar2022} and in linear perturbation theory by \citet{monari2015,monari2016}. These studies focused on moments of the DF rather than the DF itself. And though the latter treated the full 3D geometry of the disc, it didn't include self-gravity of the perturbations. Our treatment has the advantages of analytic methods while still including self-gravity. The price we pay is that the treatment of epicyclic motion and shear are only approximate. 

An outline of the paper is as follows. In Section 2, we present the formalism for calculating the response of a disc to external excitations within a shearing box. In Section 3, we consider the case where a single wave with a well-defined wave vector is excited impulsively. In particular, we study perturbations that generate either breathing waves or bending waves. In Section 4, we compute the stationary response of the disc to a co-rotating mass. We discuss the implications of our results, limitations of the shearing box approximation, and avenues for extending this work in Section 5. We conclude with a summary of our results in Section 6.

\section{Shearing box equations}

\subsection{particle orbits}

Shearing box coordinates are a local Cartesian approximation to cylindrical coordinates in a rotating frame. They were devised by \citet{hill1878} to study the three-body problem and used in early studies of galactic dynamics by \citet{goldreich1965}, JT66, and \citet{goldreich1978}. A more recent discussion of the shearing sheet in the context of galactic dynamics and swing amplification can be found in \citet{Fuchs2001}. Here we follow the pedagogical flow and notation of \citet{binney2020} (hereafter B20) who provided a particularly clear and accessible treatment of the JT66 formalism for the shearing sheet.

Let $(R,\,\phi,\,z)$ be inertial cylindrical coordinates for a rotating stellar disc and consider a patch of the disc centered on $R=R_0$, $\phi = \Omega t$ and $z=0$ where $\Omega$ is the angular frequency of a circular orbit at $R=R_0$. The shearing box coordinates are $x=R-R_0$, $y=R_0(\phi - \Omega t)$, and $z$ and the Lagrangian is given by
\begin{equation}
    {\cal L} = \frac{1}{2}\left (\dot{x}^2 + \left
    (1+\frac{x}{R_0}\right )^2\left (\Omega R_0 +
    \dot{y}\right )^2+
    \dot{z}^2\right ) - \Phi_0(x,\,z).
\end{equation}
Formally, we assume ${\bf x}\ll R_0$ and $\dot{\bf x}\ll R_0\Omega$ though these inequalities are only marginally satisfied for the size of the patch that we will consider.

The relationships between $\dot{\bf x}$ and the conjugate momenta ${\bf p}$ are given by
\begin{equation}
    \dot{x} = p_x \equiv u,
\end{equation}
\begin{equation}
    \dot{y} = \frac{p_y}{\left (1 + x/R_0\right )^2} - R_0\Omega
    \equiv v - 2Ax,
\end{equation}
\noindent and
\begin{equation}
    \dot{z} = p_Z \equiv w.
\end{equation}
The quantities $(u,v,w)$ correspond to the radial, azimuthal, and vertical components of a particle's velocity relative to the local circular orbit and are thus the shearing box analogues to the $(U,V,W)$ velocity components often used to study the dynamics of the solar neighborhood. Since the potential is independent of $y$, $p_y$ is a constant of motion. It is however, ${\cal O}(R_0\Omega)$. Following B20, we introduce $\Delta_y \equiv p_y - R_0\Omega$, which is the same order as $\dot{\bf x}$.

The Hamiltonian is given by
\begin{equation}
    H = \frac{1}{2}\left (p_x^2 + \frac{p_y^2}{\left (1 +
    x/R_0\right )^2} + p_z^2\right ) -\Omega R_0 p_y + \Phi_0(x,\,z).
\end{equation}
We assume that the potential is additively separable in $x$ and $z$ and write
\begin{equation}
    \Phi_0(x,\,z) = \xi(x) + \chi(z).
\end{equation}
Expanding $\xi$ in a Taylor series about $x=0$ we find
\begin{equation}
    \xi(x) = R_0\Omega^2x + \frac{1}{2} 
    \left (\Omega^2 - 4A\Omega\right )x^2,
\end{equation}
where $A$ is Oort's first constant and, without loss of generality, we take $\Phi({\bf x})=0$.

To quadratic order in small quantities, The Hamiltonian can be written as
\begin{equation}
    H = H_x(p_x,x) + H_y(\Delta_y) + H_z(z,p_z) +\mbox{constant}
\end{equation}
where $H_x \equiv \frac{1}{2}\left (p_x^2 + \kappa^2\left (x -\bar{x}\right )^2\right )$ and $H_z = \frac{1}{2}p_z^2 + \chi(z)$ are constants of motion and $\bar{x}\equiv 2\Omega\Delta_y/\kappa^2$ is the shearing box analogue of the guiding radius. (See B20 for a more detailed calculation.) In general, a particle will move along an elliptical orbit about the point $(x,\,y) = (\bar{x},\,y_0-2A\bar{x})$ and execute anharmonic oscillations in $z$ about the mid plane. We can therefore write
\begin{equation}
    x(t) = X\cos{\theta_r}  + \bar{x}
\end{equation}
where $\theta_r = \kappa t + \theta_0$ and $X$ and $\theta_0,$
are constants.  It follows that
\begin{equation}
    p_x = \dot{x} = -\kappa X\sin{\theta_r},
\end{equation}
and
\begin{equation}
    \dot{y} = -2A\bar{x} - 2\Omega X\cos{\theta_r},
\end{equation}
or equivalently
\begin{equation}
    v = 2B\left (x-\bar{x}\right )
\end{equation}
where $B$ is Oort's second constant. In addition, we have
\begin{equation}
    y(t) = y_0 - 2A\bar{x}t - \frac{2\Omega}{\kappa}X\sin{\theta_r}.
\end{equation}

\subsection{plane wave perturbations}

We consider a density perturbation of the form
\begin{equation}
    \rho_1({\bf x},\,t) =  e^{\ikx}\tilde{\rho_1}(z,\,t)
    \label{eq:rho1}
\end{equation}
where ${\bf x}_p$ and ${\bf k}_p$ are the position vector and wavenumber in the plane of the disc. Here and throughout, the over-tilde denotes coefficient of $e^{\ikx}$. This perturbation represents a spiral wave in the plane of the disc. If the shearing box is centered on the corotation radius of the wave, then $\kx$ must be constant for particles on circular orbits and therefore
\begin{equation}
    k_x(t_0)x + k_y y_0 = k_x(t)x + k_y y(t).
\end{equation}
For circular orbits, $x$ is constant, $y(t)-y_0 = -2A(t-t_0)x$, and therefore
\begin{equation}
    k_x(t) = k_{x0} + 2k_y A (t-t_0)
\end{equation}
where $k_{x0} = k_x(t_0)$. Without loss of generality, we can set $t_0=0$. Then
\begin{equation}
  k_p = k_y\left (1 + 4A^2t^2 + \alpha^2 + 4At\alpha\right
  )^{1/2} = k_y/\beta(t)
\end{equation}
where $\alpha\equiv k_{x0}/k_y$ and $\beta\equiv k_y/k_p$. For a single mode, we can further set $k_{x0}=0$ so that $t=0$ corresponds to the time when wave crests are aligned with the $y$ axis.

For general orbits we have
\begin{align}
  {\bf k}_p\cdot {\bf x}_p & = 
    k_{x0}\left (\bar{x} +
     X\cos{\theta_r}\right ) \nonumber\\ &+ k_y\left [y_0 + 2X\left (At\cos{\theta_r} - \frac{\Omega}{\kappa}\sin{\theta_r}\right )\right ].
 \end{align}
and therefore
\begin{equation}
  {\bf k}_p\cdot {\bf x}_p|_{t'} = 
  {\bf k}_p\cdot {\bf x}_p|_t +\psi(t')-\psi(t).
\end{equation}
where
\begin{equation}
  \psi(t) = k_{x0}X\cos{\theta_r} + 2k_yX\left (At\cos{\theta_r} -\left (\Omega/\kappa\right )\sin{\theta_r}\right).
\end{equation}

\subsection{equilibrium model}
      
By Jeans theorem, $f_0$ can be written as a function of the integrals of motion $H_x$, $\Delta_y$, and $H_z$ \citep{BT2008}. Here, we assume that it is independent of $\Delta_y$ and separable in $H_x$ and $H_z$. Following B20, we further assume that the in-plane factor of the DF is given by the Maxwell-Boltzmann distribution. For the vertical factor, we use the DF for the lowered isothermal plane \citep{weinberg1991}. Putting these together, we have
\begin{equation}
    f_0({\bf x},\,{\bf p}) = \frac{\Omega\Sigma_0}{\left (2\pi\right )^{3/2}\kappa z_0\sigma_x^2\sigma_z}
      e^{-H_x/\sigma_x^2}F_z(H_z) 
\end{equation}
where $\Sigma_0$ is the surface density, $z_0\equiv\sigma_z^2/\pi G \Sigma_0$ is the characteristic thickness of the system, and
\begin{equation}
    F_z(H_z) = \begin{cases} N_z 
    \left (e^{-H_z/\sigma_z^2} - e^{-E_0/\sigma_z^2}\right ) & 0 < H_z < E_0\\
    0 & \mbox{otherwise}.
    \label{eq:vdf0}
\end{cases}
\end{equation}
The constant $N_z$ is defined so that 
\begin{equation}
    \frac{1}{\sqrt{8\pi}\sigma_z z_0}\int dz dp_z F_z(H_z) = 1
\end{equation}
and
\begin{equation}
    \Sigma_0 = \int d^3{\bf p}\,dz f_0({\bf x},\,{\bf p}).
\end{equation}
Note that our definition of the in-plane DF differs from the one in B20 since we use $p_y$ as the azimuthal velocity coordinate rather than $v$. The vertical potential $\Phi_z$ and density are determined by solving Poisson's equation
\begin{equation}
    \frac{d^2\Phi_z}{d^2 z} = 4\pi G\rho_0(z)
\end{equation}
where
\begin{align}
    \rho_0(z) & = \int d^3{\bf p} f_0 \\ & = \frac{N_z\Sigma_0}{2z_0}
    \left (e^{-\chi_z(z)/\sigma_z^2}\,{\rm erf}(t) -
    \frac{2}{\sqrt{\pi}}t 
    e^{-E_0/\sigma_z^2}\right ) 
\end{align}
with $t = \left (E_0-\chi(z)\right )^{1/2}/\sigma_z$. Note that
for a self-consistent model, that is, one where the potential is determined entirely from the disc itself, we recover the result for the isothermal plane in the limit $E_0\to \infty$:
\begin{equation}
    \rho(z) = \frac{\Sigma_0}{2z_0}{\rm sech}^2(z/z_0)
\end{equation}
\begin{equation}
    \chi(z) = 2\sigma_z^2 \log{{\rm cosh}(z/z_0)}
\end{equation}
and $N_z=1$ \citep{spitzer1942, camm1950}. The lowered isothermal plane is the one-dimensional analogue of the lowered isothermal sphere or King model \citep{king1966} and provides a model in which the density is identically zero for $|z|$ greater than some finite truncation length.

\subsection{gravitational potential}
  
The contribution to the gravitational potential from the density perturbation in equation \ref{eq:rho1} can be written
\begin{equation}
    \Phi_1({\bf x},\,t) = e^{\ikx}\tpz (z,t)
\end{equation}
where
\begin{equation}
    \frac{\partial^2 \tpz}{\partial z^2} - k_p^2 \tpz = 4\pi G\tilde{\rho}_1(z,\,t)~.
\end{equation}
The solution is given by 
\begin{equation}
    \tpz(z,t) = -\frac{2\pi G}{k_p} P(z,t)\label{eq:potential}
\end{equation}
where the Green's function integral
\begin{equation}
P(z,t) = \int_{-\infty}^\infty
    \tilde{\rho}_1(\zeta,t)e^{-k_p|z-\zeta|}\,d\zeta.
    \label{eq:potentialGF}
\end{equation}
has dimensions of surface density. The result for a razor-thin disc,
\begin{equation}
    \tpz(z,t) = -\frac{2\pi G}{k_p} \Sigma_1(t)e^{-k_p|z|},
\end{equation}
is recovered by setting $\tilde{\rho}_1(\zeta,t) = \tilde{\Sigma}_1(t)\delta(\zeta)$. Similarly, the $z$-derivative of the potential is given by
\begin{equation}
      \frac{\partial \tpz}{\partial z} = 2\pi G Q(z,t)
\end{equation}
where
\begin{equation}
    Q(z,t) \equiv \int_{-\infty}^\infty\tilde{\rho}(\zeta,t) e^{-k_p|z-\zeta|}{\rm sgn}(z-\zeta) d\zeta.
    \label{eq:potentialDGF}
\end{equation}

\subsection{linearized distribution function}

In linear theory, we write $f({\bf x},\,{\bf p},\,t) = f_0(z,\,p_z) + f_1({\bf x},\,{\bf p},\,t)$ and $H = H_0 + \Phi_1({\bf x},\,t)$. The collisionless Boltzmann equation is then
\begin{equation}
    \frac{df_1}{dt} \equiv \frac{\partial f_1}{\partial t} + \left [f_1,\,H_0,\right ]  = \left [\Phi_1,\,f_0\right ]
\end{equation}
where $\left [~,~\right ]$ are the usual Poisson brackets \citep{BT2008}. This equation admits the following integral expression for $f_1$:
\begin{equation}
   f_1({\bf x},{\bf p},t) = J_p + J_z
   \label{eq:f1JpJz}
 \end{equation}
where
\begin{equation}
    J_p({\bf x},\,{\bf p})
    \equiv \int^t_{t_i}  dt'\,
    \frac{\partial\Phi_1 }{\partial {\bf x_p}}\cdot 
    \frac{\partial f_0 }{\partial {\bf p}_p}
    =i\int^t_{t_i}  dt'\,{\bf k}_p\cdot\frac{\partial f_0 }{\partial {\bf p}_p}\,\Phi_1 
\end{equation}
and
\begin{align}
    J_z({\bf x},\,{\bf p})&\equiv
          \int^t_{t_i} dt'\,\frac{\partial\Phi_1 }{\partial z}\frac{\partial f_0 }{\partial p_z }.
\end{align}
The lower bounds for the integrals represent an initial time when perturbations are first introduced while the integrands are evaluated along unperturbed orbits as given in Section 2.1. For $J_p$, we use the fact that
\begin{equation}
    \frac{\partial f_0}{\partial p_y} = 
    \frac{\partial f_0}{\partial \Delta_y} = 
    \frac{2\Omega}{\kappa^2}\frac{\partial f_0}{\partial \bar{x}}
\end{equation}
to find 
\begin{equation}
    {\bf k_p}\cdot \frac{\partial f_0}{\partial 
    {\bf p}_p} = -\frac{f_0}{\sigma_x^2} \left
    (k_xp_x - 2\Omega k_y (x-\bar{x})\right ).
\end{equation}
Since this expression is evaluated along an unperturbed orbit, we have
\begin{equation}
    \frac{\bf k_p}{k_p}\cdot \frac{\partial f_0}{\partial 
    {\bf p}_p} = \frac{f_0\beta}{\sigma_x^2}
    \left (\left (\alpha+ 2At'\right )\kappa X\sin{\theta_r} + 2\Omega X\cos{\theta_r}\right ).
\end{equation}
       
For a plane wave perturbation
\begin{equation}
    \rho_1({\bf x}',t')  =
    e^{i{\bf k}_p'\cdot {\bf x}_p'}\tilde{\rho}_1(z',\,t')= 
    e^{i{\bf k}_p\cdot {\bf x}_p}e^{i\left (\psi'-\psi\right )}\,\tilde{\rho}_1(z',t')
\end{equation}
and therefore
\begin{align}
    \tilde{J}_p({\bf x},{\bf p},\,t)  & =
    -\frac{iG\Sigma_0\Omega}{\left (2\pi\right
    )^{1/2}\kappa z_0\sigma_x^4\sigma_z}
    e^{-H_x/\sigma_x^2} F_z(H_z) \nonumber\\
    & \times \int^t_{t_i} dt'\,\beta' e^{i(\psi'-\psi)} P(z',t')\nonumber\\
    & \times\left (\left (\alpha + 2At'\right )\kappa\, X\sin{\theta_r} +
    2\Omega\,X\cos{\theta_r}\right ).
\end{align}
We recover equation 38 of B20 by setting $\alpha=0$, $\tilde{\rho}_1 = \tilde{\Sigma}_1\delta(z)$, and $F_z = \sqrt{8\pi}z_0\sigma_z\delta(z)\delta(p_z)$ and then integrating
over $z$ and $w$.

A similar calculation leads to the following expression for $J_z$:
\begin{align}
    \tilde{J}_z ({\bf x},{\bf p},\,t)  
    & = -\frac{G\Sigma_0\Omega N_z}{\sqrt{2\pi}\kappa\sigma_x^2 \sigma_z^3}
    e^{-H_x/\sigma_x^2}e^{-H_z/\sigma_z^2}\nonumber\\
    & \times \int^t_{t_i} dt'\,w' e^{i\left (\psi'-\psi
    \right )}Q(z',t').
\end{align}

\subsection{vertical phase space DF}

We obtain the DF in the $z-w$ plane by integrating $f_1$ over ${\bf p}_p$. Following equation \ref{eq:f1JpJz} we write
\begin{equation}
    f_{1z}(z,\,p_z) = e^{\ikx}\left (\tilde{\cal J}_p + \tilde{\cal J}_z\right )
    \label{eq:vdf}
\end{equation}
where $\tilde{\cal J}_{p,z}\equiv \int d^2 {\bf p}_p\tilde{ J}_{p,z}$. To carry out the integral we use the following change of variables from B20. Since $dp_y = d\Delta_y = \left (\kappa^2/2\Omega\right )d\bar{x}$, we have $dp_x dp_y = \left (\kappa/2\Omega\right )dU_x' dU_y'$ where
\begin{equation}
    U_x' = p_x = -\kappa X\sin{\theta_r}
    \label{eq:Uxp}
\end{equation}
and
\begin{equation}
    U_y' = \kappa\left (x-\bar{x}\right ) = \kappa X\cos{\theta_r}.
    \label{eq:Uyp}
\end{equation}
In the $(U_x',\,U_y')$ system, $\theta_r$ is a polar angle. We can therefore use $(U_x,\,U_y)$ coordinates that are obtained from $(U_x',\,U_y')$ by rotating $\theta_r$ into $\theta_0$. The result is
\begin{align}
    \tilde{\cal J}_p (z,w,t)  & = -\frac{i G\Sigma_0 F_z(H_z)}{\sqrt{2\pi} z_0\sigma_x^4\sigma_z} 
    \int^t_{t_i} dt'\beta' P(z',t') \nonumber\\
    & \times\int d^2U {\bf c}\cdot {\bf U}e^{-U^2/2\sigma_x^2 + 2i{\bf b}\cdot {\bf U}}
\end{align}
where the vectors ${\bf b}$ and ${\bf c}$ are defined in B20 and the Appendix. The integral over ${\bf U}$ can be done analytically to yield
\begin{align}
    \tilde{\cal J}_p(z,w,t)  &= \frac{\sqrt{8\pi}
    G\Sigma_0 F_z(H_z)}{z_0\sigma_z}\\
    &\int^t_{t_i} dt'\,\beta' {\bf c}\cdot {\bf b}e^{-2\sigma_x^2 b^2}P(z',t').
    \end{align}
This expression can be written in terms of the Toomre parameter
\begin{equation}
    Q \equiv \frac{\kappa\sigma_x}{3.36 G\Sigma_0}
\end{equation}
and the critical wavenumber for axisymmetric perturbations
\begin{equation}
    k_{\rm crit} \equiv \frac{\kappa^2}{2\pi G\Sigma_0}.
\end{equation}
The result is
\begin{equation}
    \tilde{\cal J}_p(z,w,t) = \frac{\kappa F_z(H_z)}{\sqrt{8\pi}z_0\sigma_z}\int^t_{t_i} dt' K_p(t,t') P(z',t')
\end{equation}
where 
\begin{equation}
    K_p(t,t') = 4\beta' {\bf c}\cdot \hat{\bf b}
    \exp(-0.572Q^2 \hat{b}^2) 
\end{equation}
and $\hat{\bf b} = (\kappa/k_{\rm crit}){\bf b}$. This equation is an example of a Volterra integral and extends the well-known result from JT66 into the dimension perpendicular to the mid plane of the disc. The function $K_p$, which B20 referred to as the JT kernel, is the same as in the case of a razor thin disc. The difference here is that it is multiplied by the Green's function integral, which is also a function of $t'$. Thus, the effective kernel is $K_pP$.

A similar calculation leads to
\begin{equation}
    \tilde{\cal J}_z = -\frac{N_ze^{-H_z/\sigma_z^2}}{\sqrt{8\pi}z_0^2\sigma_z}\int^t_{t_i} dt' p_z' K_z(t,t')Q(z',t')
\end{equation}
where 
\begin{equation}
    K_z(t,t') = \exp(-0.572Q^2 \hat{b}^2). 
\end{equation}
The integral of this term over $z$ and $w$ is zero and therefore it doesn't contribute directly to $\Sigma_1$. In short, ${\cal J}_p$ describes the redistribution of mass in the plane of the disc while ${\cal J}_z$ describes the redistribution of mass in the $z-w$ plane. 

\subsection{physical parameters}

For definiteness we set physical quantities for our calculations as follows. We take the angular frequency of the shearing box to be $\Omega = V_c/R_0 = (230\kms) /8\,{\rm kpc} \simeq 28.8\,\kmskpc$ and Oort's first constant to be $A = \Omega/2 \simeq 14.4\,\kmskpc$. The choice of $A=-B=\Omega/2$ corresponds to a Mestel disc, which has a flat rotation curve. The epicycle frequency is then $\kappa = \sqrt{2}\Omega \simeq 41\,\kmskpc$. Following B20, we use $\kappa t/\pi$ as a dimensionless time variable when plotting the time evolution of various quantities. For reference, $\pi/\kappa \simeq 77\,{\rm Myr}$. We assume a surface density for the equilibrium system of $\Sigma_0 = 7\times 10^7\,M_\odot {\rm kpc}^{-2}$, which gives a volume density in the mid plane of $\rho_0 \simeq 0.16\left (\sigma_z/15\,\kms\right )^{-2}M_\odot\,{\rm pc}^{-3}$. With these parameters, the critical wavelength if $\lambda_{\rm crit} \simeq 8.2\,{\rm kpc}$. These values are roughly consistent with values for the Solar neighbourhood. Of course, a more realistic model would include a multi-component disc along with other contributions to the gravitational potential such as the gas disc and dark halo. 

\section{impulsive excitations}

\subsection{breathing wave excitation}

We first consider the response of the disc to a plane-wave impulsive excitation. We begin by assuming that the excitation is localized in the mid plane and symmetric in $z$. The density in equations \ref{eq:potentialGF} and \ref{eq:potentialDGF} is then given by
\begin{equation}
    \tilde{\rho}_1(z,t) = \frac{\Sigma_e}{\kappa}\delta\left (t-t_i\right )\delta(z) + \tilde{\rho}_s(z,t).
\end{equation}
where $\rho_s$ is the density perturbation in the disc itself, i.e., the self-gravity term. Since this is a single wave with well defined $k_y$, we can set $k_x(t=0) = k_{x0}=0$. In the absence of self-gravity, the vertical DF given by
\begin{align}
    & \tilde{f}_{1z}(z,w,t)  = \Sigma_e e^{-k_p|z_i|} \nonumber\\
    & ~~\times \left (K_p(t,t_i)F_{z}(H_z)-
    \frac{w_i N_z}{\kappa h_z}{\rm sgn}(z_i)K_z(t,t_i)e^{-E_z/\sigma_z^2} \right ).
\end{align}
Note that $z_i = z(t_i)$ and $w_i=w(t_i)$ are implicit functions of $z$ and $w$.

\begin{figure}
	\includegraphics[width=\columnwidth]{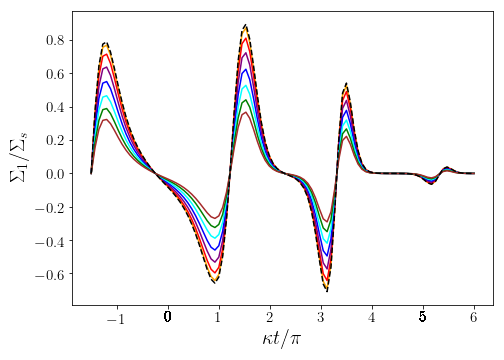}
       \caption{Amplitude of an impulsively-excited plane wave as a function of time in the absence of self-gravity. The amplitude has been normalized by the amplitude of the impluse. Shown are results for discs of various thicknesses as set by the vertical velocity dispersion where $\sigma_z = 5-30\,{\rm km\,s^{-1}}$ in steps of $5\,{\rm km\,s^{-1}}$ for orange, red, purple, cyan, green, and brown. The black curve is the result for a razor thin disk. }
  \label{fig:Sigma_noSG}
\end{figure}

\begin{figure}
	\includegraphics[width=\columnwidth]{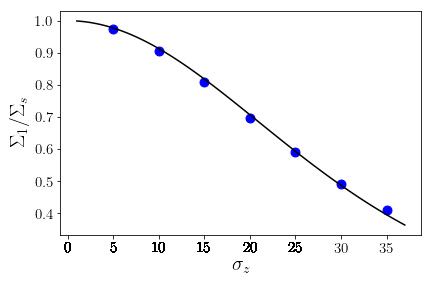}
       \caption{Reduction in amplitude of a kinematic wave as a function of $\sigma_z$. The blue points give the reduction in amplitude relative the result for the razor thin disc. The black curve is the phenomenological formula due to \citet{toomre1964} and JT66} as discussed in the text. 
  \label{fig:reduction}
\end{figure}

In Fig. \ref{fig:Sigma_noSG} we show the evolution of $\Sigma_1/\Sigma_e$ for the case when $t_i=-1.5\pi/\kappa$, $Q=1.2$, and $k_y=k_{\rm crit}/2$. The timescale between successive peaks corresponds to the period for epicyclic motions, $2\pi/\kappa$. The figure illustrates the effect the disc's thickness has on the effective kernel since, in the absence of self-gravity, $\Sigma_1/\Sigma_e$ is proportional to $K_pP$. In Fig. \ref{fig:reduction} we plot the reduction in the peak value of $\Sigma_1$ relative to what the peak value would be in a razor-thin disc. \citet{toomre1964} and JT66 suggested that one could account for finite thickness effects by multiply $K$ by $(1-\exp{\gamma})/\gamma$ where $\gamma\equiv hk_y(1 + (2At_i)^2)^{1/2}$ and $2h$ is the effective thickness of the disc. This expression can derived from equations \ref{eq:potential} and \ref{eq:potentialGF} by taking the disc is a uniform density slab of thickness $2h$. For the fit in Fig.\,\ref{fig:reduction} we use $hk_y = 0.9 z_0k_{\rm crit}/2 = 0.9\left (0.53Q\sigma_z/\sigma_x\right )^2$ where the numerical factor of $0.9$ is obtained via chi-by-eye and accounts for the fact the density in our truncated isothermal model is not uniform.

\begin{figure}
	\includegraphics[width=\columnwidth]{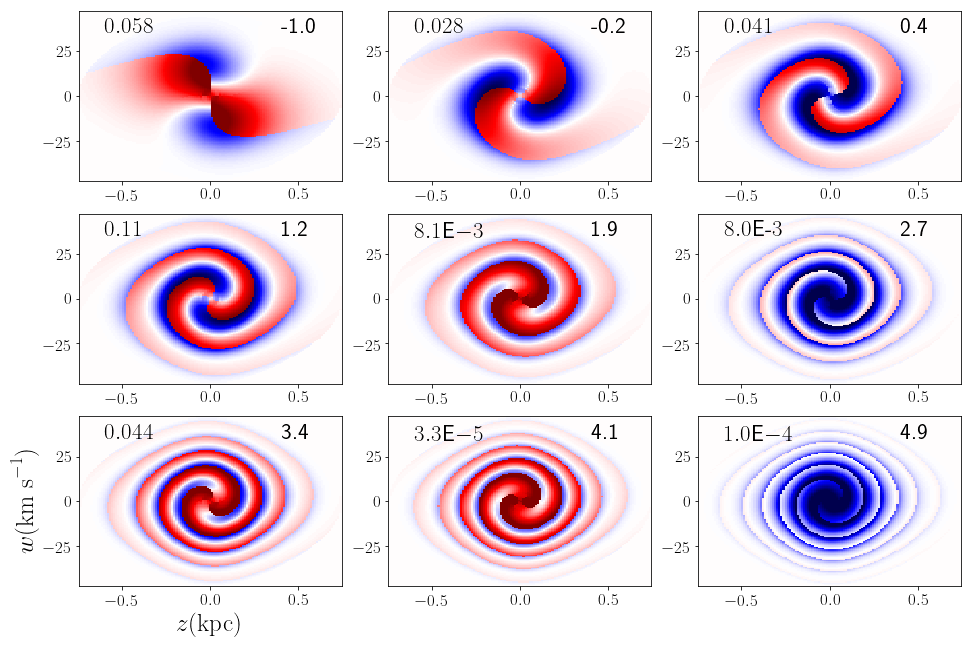}
       \caption{Vertical ($z-w$) phase space DF when self-gravity is ignored at nine different epochs for the case where $\sigma_z=15\kms$. The numbers in the upper right corners of each panel give the time in units of $\pi/\kappa$. The numbers in the upper left corners indicate the maximum value for the phase space density as indicated by the color scale in the sense that one integrates the map over $z$ and $w$ to obtain the value of $\Sigma_1/\Sigma_e$ in Fig.\ref{fig:Sigma_noSG}.}
  \label{fig:fzw_noSG}
\end{figure}

In Fig.\,\ref{fig:fzw_noSG} we show the vertical DF at nine different times for the case where $\sigma_z=15\,\kms$. The DF winds up with a pitch angle that increases linearly with time (see below). The time-dependence of the amplitude of the perturbation is consistent with what we showed in Fig. \ref{fig:Sigma_noSG}. In particular, the amplitude reaches local maxima at $\kappa t/\pi\simeq -1.2,\,1.5,\,3.5$ and local minima at $\kappa t/\pi\simeq 1,3.1$. 

\begin{figure}
	\includegraphics[width=\columnwidth]{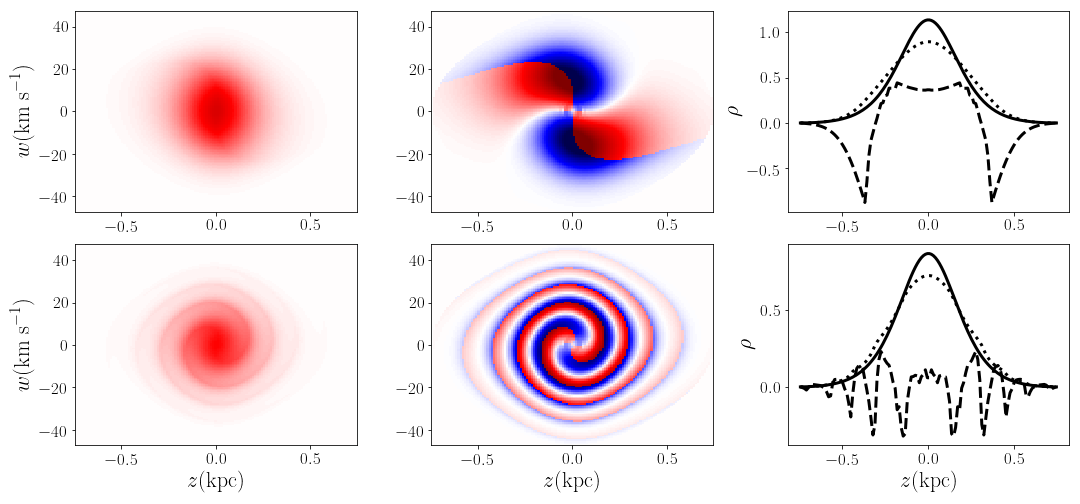}
       \caption{Separate contribution to the vertical DF from the ${\cal J}_p$ (left column) and ${\cal J}_z$ (right column). The upper row is for $\kappa t/\pi = 1$ while the lower row is for $\kappa t/\pi = 3.4$. The sum of the two contributions yields the total DF as seen in the corresponding panels of Fig.\ref{fig:fzw_noSG}. The rightmost column shows the contributions from ${\cal J}_p$ and ${\cal J}_z$ to the vertical density as dotted and dashed lines, respectively. The solid line shows the number density for the equilibrium model.}
  \label{fig:fzwdecomp.png}
\end{figure}

As discussed above, the ${\cal J}_p$ and ${\cal J}_z$ terms in equation \ref{eq:vdf} involve very different aspects of mass redistribution. The ${\cal J}_p$ term describes mass redistribution in the plane and therefore changes the local surface density. Conversely, ${\cal J}_z$ describes a redistribution of mass in the $z-w$ plane, but leaves the surface density unchanged. This difference is illustrated in Fig. \ref{fig:fzwdecomp.png} where we show separate contributions to the vertical DF from ${\cal J}_p$ and ${\cal J}_z$ at two different epochs. For $\kappa t/\pi=-1$, shortly after the disc has been excited, the contribution from ${\cal J}_p$ is proportional to $F_z(H_z)\Phi_1$, which is symmetric in $z$. This contribution changes the shape of the $z-w$ DF while preserving the $z\to -z$ and $w\to -w$ symmetries. We can therefore view it as a breathing wave along with mass redistribution in the $x-y$ plane. Over time, the pattern winds up into a two-armed phase spiral. With ${\cal J}_z$, the perturbation due to the excitation is initially proportional to $w F_z(H_z)\partial \Phi_1/\partial z$. Thus, the amplitude of the perturbation is maximal along the diagonals of the $z-w$ plane with a sign that alternates as one circles the origin. This pattern is evident in the lower middle panel of Fig. \ref{fig:fzw_noSG}. The perturbation winds up into a two armed spiral with alternating positive and negative bands. These points are further illustrated in the right-most panels where we show the density as a function of $z$. While both ${\cal J}_p$ and ${\cal J}_z$ lead to density perturbations that are symmetric in $z$, the variations with $z$ for the latter are more prominent due to the geometry of the phase space perturbation.

\begin{figure}
	\includegraphics[width=\columnwidth]{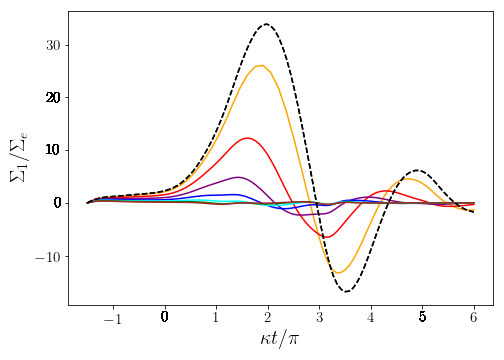}
       \caption{Surface density a function of time for the case where self-gravity is included. Physical conditions and line colors as the same as in Fig. \ref{fig:Sigma_noSG}.}
  \label{fig:Sigma_SG}
\end{figure}

In Fig.\ref{fig:Sigma_SG} we show the surface density as a function of time for the case when self-gravity, and hence swing amplification, are included. For a razor thin disc, the surface density is amplified by a factor of $\sim 35$. The amplification factor decreases with increasing disc thickness. For example, when $\sigma = 20\,\kms$, the peak is only 13\% of what it is for the razor thin case even though the kernel is only reduced by 20\%. Nevertheless, the perturbation is still amplified by a factor of $4.7$. 

\begin{figure}
	\includegraphics[width=\columnwidth]{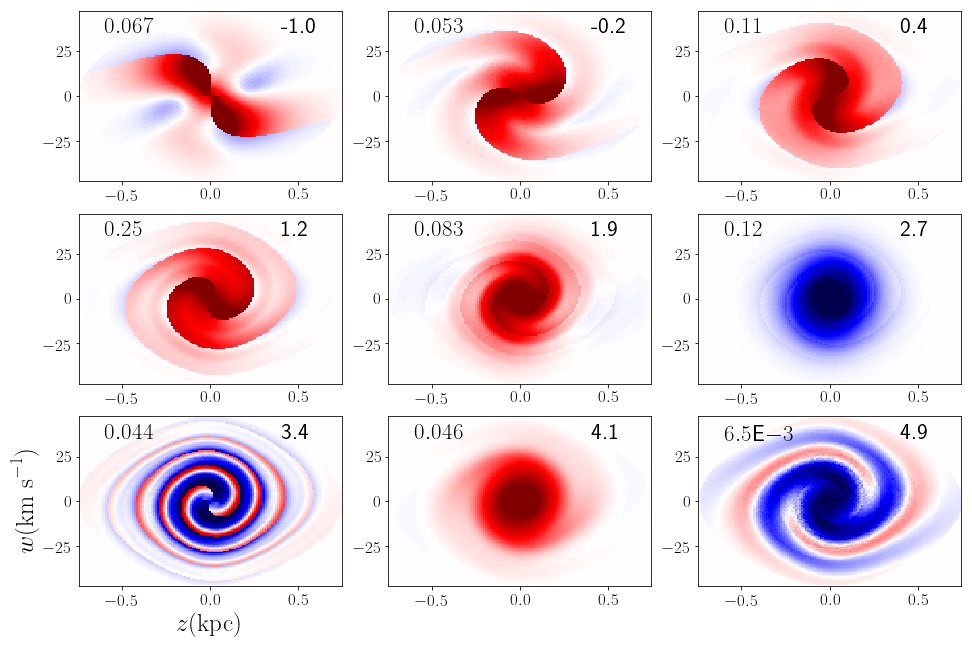}
       \caption{Vertical DF when self gravity is included. The physical conditions and epochs are the same as in Fig. \ref{fig:fzw_noSG}.}
  \label{fig:fzw_SG}
\end{figure}

In Fig.\ref{fig:fzw_SG} we plot the vertical DF for the same nine epochs as in Fig.\ref{fig:fzw_noSG}. The effects of self-gravity are striking. In the absence of self-gravity, the contributions from ${\cal J}_p$ and ${\cal J}_z$ are comparable and so we have a phase spiral combined with a modulation of the local surface density. With self-gravity, there is a rapid amplification of the surface density and compression in the $z-w$ phase space at $t \sim 1.5\pi/\kappa$. This in-plane compression of the disc sets off new vertical phase spirals so that by $t\sim 4\pi/\kappa$ the spirals are less tightly wound but stronger than in the case without self gravity. 

\begin{figure}
	\includegraphics[width=\columnwidth]{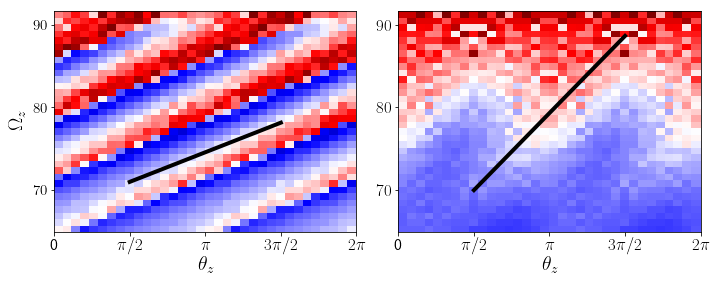}
       \caption{Vertical phase space DF in the angle-frequency ($\theta_z-\nu_z$) plane. The left panel shows the total DF for $\kappa t/\pi = 4.1$ (lower-middle panel of Fig. \ref{fig:fzw_noSG} mapped onto the $\theta_z-\nu_z$ plane. The black line-segment is the expected slope from kinematic phase mixing. It corresponds to an age of the disturbance of $t-t_i = 5.6\pi/\kappa$ and lines up with the ridges. The right panel shows the same plot for the case when self-gravity is included (lower-middle panel of Fig. \ref{fig:fzw_SG}. In this case, the black line segment corresponds to $t-t_{sw}\simeq 2.2\pi/\kappa$ where $t_{sw}$ is roughly the time of the swing-amplification peak.}
  \label{fig:theta_nu}
\end{figure}

These points are further illustrated in Fig.\ref{fig:theta_nu} where we show the DF as a function of $\Omega_z$ and  $\theta_z$. Plots of Gaia data in this space were made by \citet{LW2021, frankel2022, LW2023} and \citet{tremaine2022}. In the absence of self-gravity, the phase spiral is transformed into parallel, diagonal bands with a slope proportional to the reciprocal of the age of the spiral. With self-gravity, the bands are steeper suggesting a younger age. In fact, the slope is related to the time between the swing-amplification peak and the observation time. In addition, the bands appear to bend upward in a manner similar to what is seen in \citet{frankel2022} and \citet{tremaine2022}.

\subsection{bending wave excitation}

Our previous examples focused on breathing waves and two-armed vertical spirals. The excitation of bending waves and one-armed spirals requires an external density that is anti-symmetric in $z$. As a toy model for this process, we take the external mass distribution to be
\begin{equation}
    \tilde{\rho_e} = \frac{\Sigma_b}{\kappa\Delta^2}\delta(t-t_i)
    ze^{-z^2/2\Delta^2}~.
    \label{eq:bendpert}
\end{equation}

\begin{figure}
	\includegraphics[width=\columnwidth]{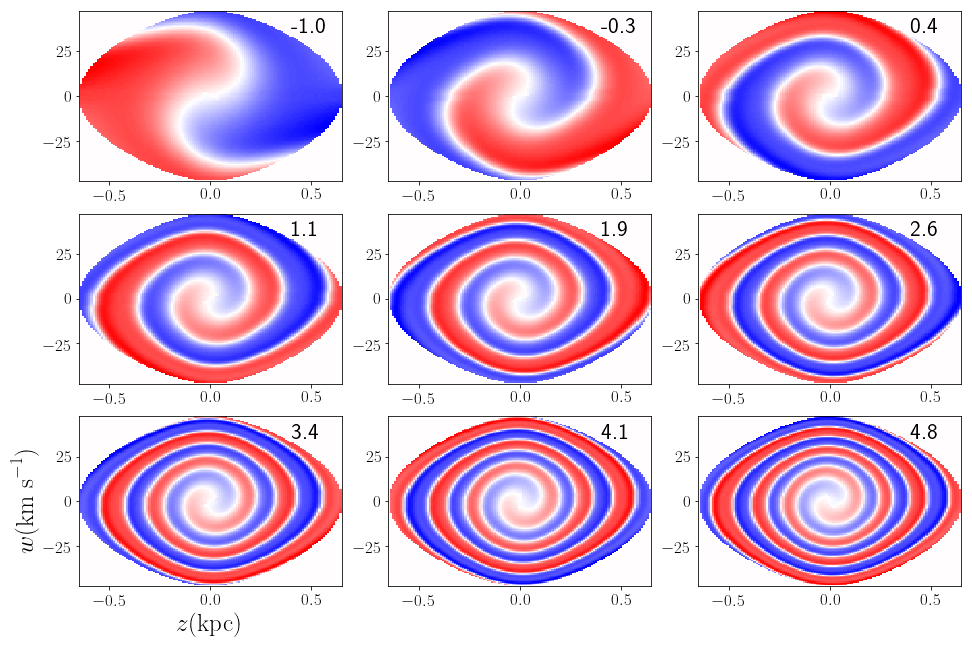}
       \caption{Vertical phase space DF when self-gravity is ignored for nine different epochs. The figure is similar to Fig.\ref{fig:fzw_noSG} except that here, the disc is excited by an anti-symmetric density given by equation \ref{eq:bendpert}. The color scale is centered on zero and its stretch is the same in each panel. Integrating the absolute value of the DF in each map over $z$ and $w$ yields $0.1\Sigma_b$ where $\Sigma_b$ is the amplitude of the external excitation in equation \ref{eq:bendpert}.}
  \label{fig:fzw_bend_noSG}
\end{figure}
\begin{figure}
	\includegraphics[width=\columnwidth]{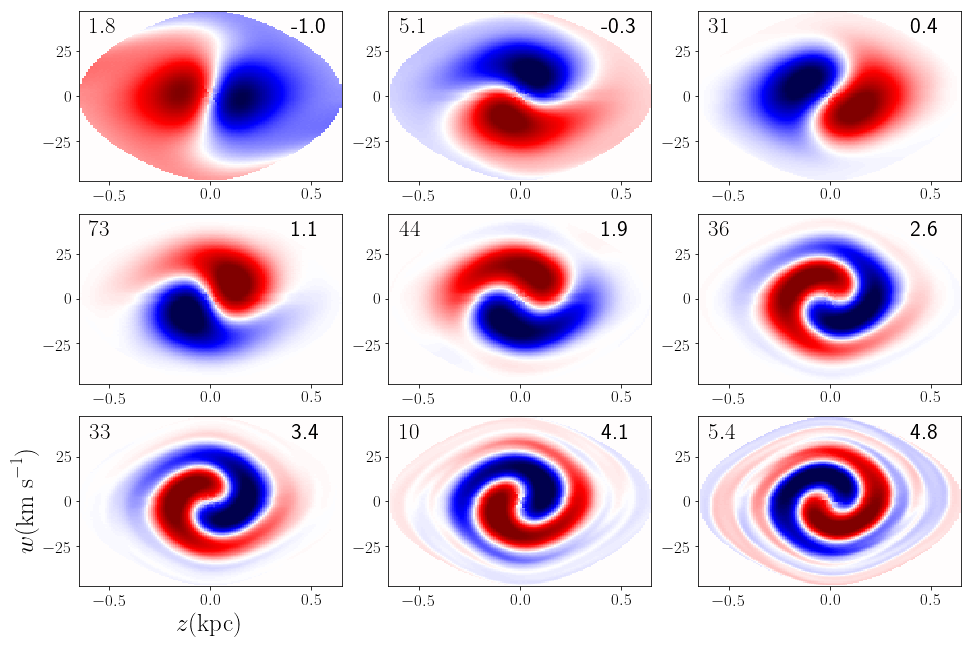}
       \caption{Vertical DF for a bending perturbation when self-gravity is included. In this case, the number in the upper left corner of each panel gives the amplification factor relative to the case where self-gravity is ignored (Fig.\,\ref{fig:fzw_bend_noSG}).} 
  \label{fig:fzw_bend_SG}
\end{figure}

The evolution of the system in the absence of self gravity is shown in Fig.\,\ref{fig:fzw_bend_noSG}. The initial anti-symmetric perturbation winds up just as one expects. On the other hand, when self gravity is included (Fig.\,\ref{fig:fzw_bend_SG}) the DF is amplified during the early stages of evolution and the rate at which the system phase mixes is significantly slower. Without self gravity the system undergoes $\sim 3$ phase wrappings by $t \simeq 5\pi/\kappa$ as compared with just a single phase wrapping in the self-gravitating case.

The importance of self gravity for an anti-symmetric disturbance may seem surprising since the total surface density is identically zero. However, one can think of the external density $\rho_e$ as the combination of two parallel discs, one with positive surface density and the other with negative surface density. Since the time scale for swing amplification is comparable to the time scale for mixing between the upper and lower components of the disc, it is able to amplify these components separately before mixing takes hold. Only after one or two wrappings is achieved does phase mixing proceed at the rate expected from pure kinematics.

\section{Excitation of the disc by a cloud}

Next, we consider the disc's response to a massive cloud on a circular orbit, that is, a cloud at rest in the shearing box. Following JT66 and B20, we decompose the density of the cloud into Fourier modes and compute the response of the disc from each mode using the formalism developed in the previous section. We assume that the perturbing mass is a Gaussian in ${\bf x}$,
\begin{equation}
\rho_e({\bf x}) = \frac{M}{\left (2\pi\right )^{3/2}\Delta^3}
e^{-|{\bf x}|^2/2\Delta^2}~,
\end{equation}
where $\Delta$ and $M$ are the size and mass of the cloud, respectively. The ${\bf x}_p$ Fourier transform is then
\begin{equation}
\tilde\rho_e({\bf k}_p,\,z) = \frac{M}{\sqrt{2\pi} \Delta} e^{-\Delta^2 |{\bf k}_p|^2/2}e^{-z^2/2\Delta^2}~.
\end{equation}
As discussed in JT66 and B20, each Fourier mode evolves according to the equation developed in the previous section. We can therefore replace $\tilde{\rho}_1(\xi,\,t')$ in equations \ref{eq:potentialGF} and \ref{eq:potentialDGF} with
$\tilde{\rho_e} + \tilde{\mu}_s\left ({\bf k}_p;\zeta,t'\right )$. Note that while $\tilde{\mu}_s$ is treated like $\tilde{\rho}_s$ in our plane-wave calculations, here it has dimensions of mass per unit length. The volume density is calculated via an inverse Fourier transform, that is, an integral over $k_x$ and $k_y$. Likewise, an inverse Fourier transform is required to get $f_{1z}$ as a function of $x$ and $y$. 

\begin{figure}
	\includegraphics[width=\columnwidth]{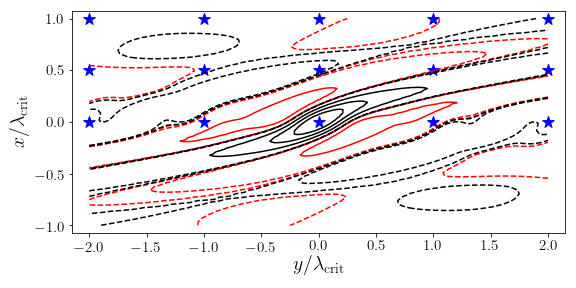}
       \caption{Surface density from a massive cloud that is on a circular orbit at ${\bf x}=0$. The contours, in units $M/{\rm kpc^2}$, are as follows: solid black -- $0.2, 0.4, 0.6$; dashed black -- $0.001, 0.01$; solid red -- $-0.2$; dashed red -- $-0.01, 0.001$.}
  \label{fig:cloudxy}
\end{figure}

In Fig.\ref{fig:cloudxy} we show the surface density perturbation generated by the cloud for the case when $\Delta = 0.05\lambda_{\rm crit}$ and $Q=1.2$. The features of the response have been discussed at length in JT66, \citet{Fuchs2001}, and B20. The ridge that runs from the lower left to the upper right arises from swing-amplified trailing waves that originated as leading waves and roughly follows the line $x = y/2At_{sw}$ where $t_{sw}\simeq \pi/\kappa$ is the time of the primary swing-amplification peak. The structure is a stationary disturbance in the disc though individual stars are continuously passing through it.

\begin{figure}
    \includegraphics[width=\columnwidth]{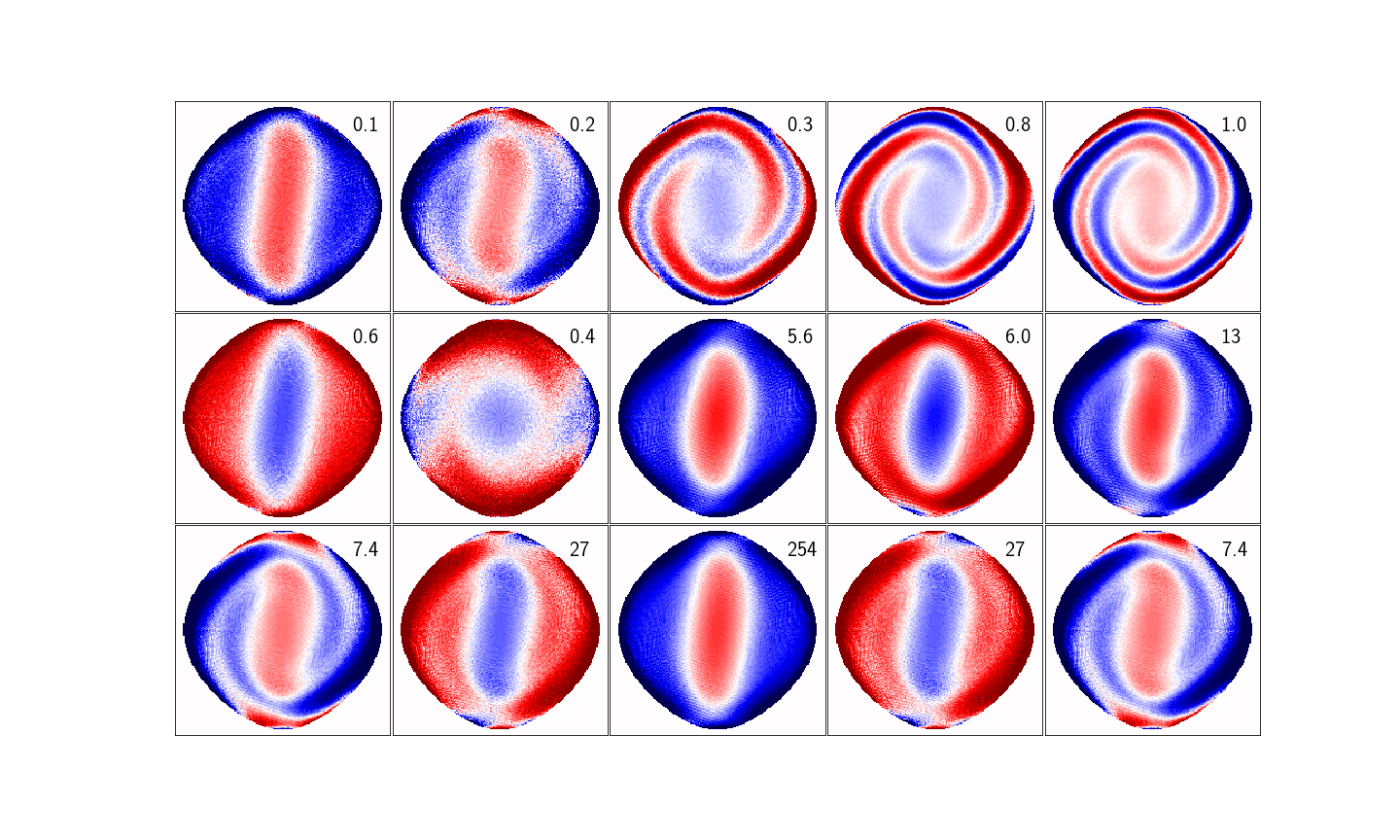}
       \caption{Thumbnail images of the fractional residuals of the $z-w$ DF across the $x-y$ plane. The positions of the images correspond to the blue stars in Fig. \ref{fig:cloudxy}. Images for the lower half of Fig.\ref{fig:cloudxy} are obtained by letting ${\bf x}_p\to -{\bf x}_p$. The fractional residual is found by dividing $f_{1z}$ by the equilibrium DF.
       The numbers in the upper right corners of each panel indicate the maximum residual as a percentage under the assumption that $M=10^9\,M_\odot$.}
  \label{fig:zwthumbnails}
\end{figure}

In Fig.\ref{fig:zwthumbnails} we show the vertical DF at the 15 positions across the disc indicated by blue stars in Fig.\ref{fig:cloudxy}. These DFs are calculated on a $200\times 200$ grid in the $z-w$ plane and then smoothed using the \textsc{Scipy} routine \textsc{ndimage.gaussian\_filter} with \textsc{sigma}$=8$. We then divide the DFs by the equilibrium DF to obtain a fractional residual map. The numbers in the upper right corners of each panel indicate the residual as a percentage assuming $M=10^9\,M_\odot$. Evidently, the pattern of perturbations in the $z-w$ plane is strongly dependent on the position within the mid plane. The phase spirals are most prominent between one and two times $\lambda_{\rm crit}$ from the perturbing mass and close to the wake produced by swing amplification. We might have anticipated this from Fig.\,\ref{fig:fzw_SG} where we found that the spirals arose soon after peaks formed in surface density. Elsewhere in the $x-y$ the perturbation takes the form of a breathing wave.

\section{Discussion}

\citet{toomre1981} described swing amplification as a conspiracy between shear, epicyclic motion or shaking, and self-gravity. The formalism presented in Section 2 allows us to study the actions of an additional conspirator: phase mixing in the dimension perpendicular to the mid plane. Phase mixing regulates swing amplification since the effects of self-gravity are diminished once the system has undergone one or two phase wrappings in the $z-w$ plane. Our formalism also allows us to explicitly account for the reduction in self-gravity due to the finite thickness of the disc. We are therefore able to demonstrate the validity of the phenomenological formula from \citet{toomre1964} and JT66.

The key takeaways from Section 3 and 4 are that swing amplification can enhance phase spirals by amplifying a disturbance in the $z-w$ plane before phase mixing takes hold. Moreover, stationary phase spirals can form in the wake of a co-rotating mass. In the usual picture of phase mixing, individual stars follow the ridges and troughs of the vertical DF as it winds up. Here, individual stars pass through the spiral in the same way that stars pass in and out of the spiral arms of a disc galaxy. Taken together, these results call into question the simple picture of the Gaia phase spirals where their shapes are determined by kinematic phase mixing.

There are obvious improvements and extensions that we can make to the calculations presented in this paper. First, external components to the background potential can be included to account for a thin gas disc and extended dark halo. Doing so will change the vertical potential and hence rate at which disturbances in the $z-w$ plane wind up while reducing the effect of self-gravity. (See, figure 2 of \citet{tremaine2022}). Second, we might consider more realistic scenarios for exciting the disc such as a passing satellite galaxy or dark matter sub halo, as in Section 6 of B20. One expects that the response of the disc to a passing satellite will be intermediate to the response from an impulsive excitation and a stationary cloud. In addition, one might add intermediate-scale masses to stir up the system as a means of testing the diffusion hypothesis in \citet{tremaine2022}.

It is worth commenting on the computational complexity of our calculations. In the shearing sheet, the computational complexity for the response to a single wave of definite $k_y$ is ${\cal O}(N_t^2)$ where $N_t$, the number of time steps in the Volterra integral, is ${\cal O}(10^2)$ for the calculations presented in this paper. The computational complexity for the response to a general, time-dependent excitation will then be $N_k^2N_t^2$ where $N_k$ is the number of points for each dimension of the ${\bf x}_p-{\bf k}_p$ Fourier transform. However, for a stationary perturbation, such as the co-moving cloud, one can use the time-invariance on the solution to replace the $k_x$ integral with an integral over $t$. The computational complexity is therefore reduced to $N_kN_t^2$ (B20). In our calculations, the complexity is increased by a factor of $N_z^2N_w$ where $N_z$ and $N_w$ and the number of grid points in the $z$ and $w$ directions, respectively. Calculations for the co-rotating cloud are therefore $N_z^2N_wN_kN_t^2$ and can be done in one to several hours on an 8 processor 3.2 GHz machine. The computation time for a general time-dependent perturbation will be on the order of one to a few days, long but not prohibitive. The computation might be improved by using angle action variables in place of $z$ and $w$ as in \citet{banik2022}. Alternatively, one might resort to N-body simulations in the shearing box \citep{fuchs2005}.

Of course, in the end, the shearing box is not a perfect substitute for a rotating disc. The critical wavelength is $2-3$ times the exponential scale length of the disc. This means that the surface density of a realistic disc varies considerably on scales considered in this paper and the separability assumption for the potential is surely suspect.

\section{Conclusions}

In this work, we presented a formalism to study the response of a small local patch of a stellar disc to an external perturbation within the framework of the shearing box approximation. It extended the shearing sheet formalism of JT66 in the dimension perpendicular to the disc and allowed us to examine what happens within the disc as it responds to an external perturbation.

In general any disturbance in the vertical structure of a disc leads to $z-w$ spirals as the disturbance undergoes phase mixing. The main result of this paper is that self-gravity can amplify disturbances in the disc before phase mixing takes hold. This amplification is strongest for leading waves as they swing into trailing ones. Perhaps unexpectedly, the process also works with bending waves. Finally, it is possible to set up stationary phase spirals in the wake of a co-rotating mass. 

A complete understanding of the Gaia phase spirals is still lacking. Investigations that range from test particles in one dimension to fully self-consistent N-body simulations have failed to reproduce the structures found in the data in all their complexity. The shearing box calculations presented in this work provide an intermediate approach since they include self-gravity and an approximate form of epicyclic motion and differential rotation or shear. Thus, despite their obvious limitations, they allow one to explore
the effects of self-gravity and the interplay between in-plane and vertical dynamics.

\section*{acknowledgements}

We acknowledge the financial support of the Natural Sciences and Engineering Research Council of Canada. 

\appendix

\section{Kernel definitions}

With the change of variables described in Section 2.6 we have $H_x = U^2$. Furthermore, we can use standard trigonometric identities to write

\begin{equation}
    \psi(t')-\psi(t) = 2{\bf b}\cdot {\bf U}
\end{equation}
where
\begin{equation}
    b_x(t,t') \equiv \frac{k_x}{\kappa} \left (A(t'S'-tS) +
    \frac{\alpha}{2}\left (S'-S\right ) +
    (\Omega/\kappa)(C'-C)\right ) 
\end{equation}
  and        
\begin{equation}
    b_y(t,t') \equiv \frac{k_y}{\kappa} \left (A(t'C'-tC) +
    \frac{\alpha}{2}\left (C'-C\right ) -
    (\Omega/\kappa)(S'-S)\right ) .
\end{equation}
Likewise, we have
\begin{equation}
    \left (\alpha + 2At'\right )\kappa X\sin{\theta_r} + 2\Omega X\cos{\theta_r} = 
    2{\bf c}\cdot {\bf U}
\end{equation}
where
\begin{equation}
    c_x' \equiv -\left (\frac{\alpha}{2} + At'\right )C' +
    \frac{\Omega}{\kappa}S' 
\end{equation}
and
\begin{equation}
    c_y' \equiv \left (\frac{\alpha}{2} + At'\right )S' + \frac{\Omega}{\kappa}C' ~.
\end{equation}

\section*{Data Availability}

The data underlying this article were generated by numerical calculations using original \textsc{Python} code written by the author. The code incorporated routines from \textsc{NumPy} \citep{harris2020} and \textsc{SciPy} \citep{virtanen2020}. The data for the figures and the code will be shared on reasonable request to the author.

\bibliographystyle{mnras}
\bibliography{ShearingBox}

\begin{thebibliography}{}
\makeatletter
\relax
\def\mn@urlcharsother{\let\do\@makeother \do\$\do\&\do\#\do\^\do\_\do\%\do\~}
\def\mn@doi{\begingroup\mn@urlcharsother \@ifnextchar [ {\mn@doi@}
  {\mn@doi@[]}}
\def\mn@doi@[#1]#2{\def\@tempa{#1}\ifx\@tempa\@empty \href
  {http://dx.doi.org/#2} {doi:#2}\else \href {http://dx.doi.org/#2} {#1}\fi
  \endgroup}
\def\mn@eprint#1#2{\mn@eprint@#1:#2::\@nil}
\def\mn@eprint@arXiv#1{\href {http://arxiv.org/abs/#1} {{\tt arXiv:#1}}}
\def\mn@eprint@dblp#1{\href {http://dblp.uni-trier.de/rec/bibtex/#1.xml}
  {dblp:#1}}
\def\mn@eprint@#1:#2:#3:#4\@nil{\def\@tempa {#1}\def\@tempb {#2}\def\@tempc
  {#3}\ifx \@tempc \@empty \let \@tempc \@tempb \let \@tempb \@tempa \fi \ifx
  \@tempb \@empty \def\@tempb {arXiv}\fi \@ifundefined
  {mn@eprint@\@tempb}{\@tempb:\@tempc}{\expandafter \expandafter \csname
  mn@eprint@\@tempb\endcsname \expandafter{\@tempc}}}

\bibitem[\protect\citeauthoryear{{Antoja} et~al.,}{{Antoja}
  et~al.}{2018}]{antoja2018}
{Antoja} T.,  et~al., 2018, \mn@doi [\nat] {10.1038/s41586-018-0510-7}, \href
  {https://ui.adsabs.harvard.edu/abs/2018Natur.561..360A} {561, 360}

\bibitem[\protect\citeauthoryear{{Antoja}, {Ramos}, {Garc{\'\i}a-Conde},
  {Bernet}, {Laporte}  \& {Katz}}{{Antoja} et~al.}{2022}]{antoja2022}
{Antoja} T.,  {Ramos} P.,  {Garc{\'\i}a-Conde} B.,  {Bernet} M.,  {Laporte}
  C.~F.~P.,   {Katz} D.,  2022, \mn@doi [arXiv e-prints]
  {10.48550/arXiv.2212.11987}, \href
  {https://ui.adsabs.harvard.edu/abs/2022arXiv221211987A} {p. arXiv:2212.11987}

\bibitem[\protect\citeauthoryear{{Araki}}{{Araki}}{1985}]{araki1985}
{Araki} S.,  1985, PhD Thesis, Massechusetts Institute of Technology

\bibitem[\protect\citeauthoryear{{Banik}, {Weinberg}  \& {van den
  Bosch}}{{Banik} et~al.}{2022}]{banik2022}
{Banik} U.,  {Weinberg} M.~D.,   {van den Bosch} F.~C.,  2022, \mn@doi [\apj]
  {10.3847/1538-4357/ac7ff9}, \href
  {https://ui.adsabs.harvard.edu/abs/2022ApJ...935..135B} {935, 135}

\bibitem[\protect\citeauthoryear{{Bennett} \& {Bovy}}{{Bennett} \&
  {Bovy}}{2019}]{bennett2019}
{Bennett} M.,  {Bovy} J.,  2019, \mn@doi [\mnras] {10.1093/mnras/sty2813},
  \href {https://ui.adsabs.harvard.edu/abs/2019MNRAS.482.1417B} {482, 1417}

\bibitem[\protect\citeauthoryear{{Bennett} \& {Bovy}}{{Bennett} \&
  {Bovy}}{2021}]{bennett2021}
{Bennett} M.,  {Bovy} J.,  2021, \mn@doi [\mnras] {10.1093/mnras/stab524},
  \href {https://ui.adsabs.harvard.edu/abs/2021MNRAS.503..376B} {503, 376}

\bibitem[\protect\citeauthoryear{{Bennett}, {Bovy}  \& {Hunt}}{{Bennett}
  et~al.}{2022}]{bennett2022}
{Bennett} M.,  {Bovy} J.,   {Hunt} J. A.~S.,  2022, \mn@doi [\apj]
  {10.3847/1538-4357/ac5021}, \href
  {https://ui.adsabs.harvard.edu/abs/2022ApJ...927..131B} {927, 131}

\bibitem[\protect\citeauthoryear{{Binney}}{{Binney}}{2020}]{binney2020}
{Binney} J.,  2020, \mn@doi [\mnras] {10.1093/mnras/staa1485}, \href
  {https://ui.adsabs.harvard.edu/abs/2020MNRAS.496..767B} {496, 767}

\bibitem[\protect\citeauthoryear{{Binney} \& {Sch{\"o}nrich}}{{Binney} \&
  {Sch{\"o}nrich}}{2018}]{binney2018}
{Binney} J.,  {Sch{\"o}nrich} R.,  2018, \mn@doi [\mnras]
  {10.1093/mnras/sty2378}, \href
  {https://ui.adsabs.harvard.edu/abs/2018MNRAS.481.1501B} {481, 1501}

\bibitem[\protect\citeauthoryear{{Binney} \& {Tremaine}}{{Binney} \&
  {Tremaine}}{2008}]{BT2008}
{Binney} J.,  {Tremaine} S.,  2008, {Galactic Dynamics: Second Edition}

\bibitem[\protect\citeauthoryear{{Bland-Hawthorn} et~al.,}{{Bland-Hawthorn}
  et~al.}{2019}]{blandhawthorn2019}
{Bland-Hawthorn} J.,  et~al., 2019, \mn@doi [\mnras] {10.1093/mnras/stz217},
  \href {https://ui.adsabs.harvard.edu/abs/2019MNRAS.486.1167B} {486, 1167}

\bibitem[\protect\citeauthoryear{{Camm}}{{Camm}}{1950}]{camm1950}
{Camm} G.~L.,  1950, \mnras, 110

\bibitem[\protect\citeauthoryear{{Carlin} et~al.,}{{Carlin}
  et~al.}{2013}]{carlin2013}
{Carlin} J.~L.,  et~al., 2013, \mn@doi [\apjl] {10.1088/2041-8205/777/1/L5},
  \href {https://ui.adsabs.harvard.edu/abs/2013ApJ...777L...5C} {777, L5}

\bibitem[\protect\citeauthoryear{{Darling} \& {Widrow}}{{Darling} \&
  {Widrow}}{2019}]{darling2019a}
{Darling} K.,  {Widrow} L.~M.,  2019, \mn@doi [\mnras] {10.1093/mnras/sty3508},
  \href {https://ui.adsabs.harvard.edu/abs/2019MNRAS.484.1050D} {484, 1050}

\bibitem[\protect\citeauthoryear{{Debattista}}{{Debattista}}{2014}]{debattista2014}
{Debattista} V.~P.,  2014, \mn@doi [\mnras] {10.1093/mnrasl/slu069}, \href
  {https://ui.adsabs.harvard.edu/abs/2014MNRAS.443L...1D} {443, L1}

\bibitem[\protect\citeauthoryear{{Frankel}, {Bovy}, {Tremaine}  \&
  {Hogg}}{{Frankel} et~al.}{2022}]{frankel2022}
{Frankel} N.,  {Bovy} J.,  {Tremaine} S.,   {Hogg} D.~W.,  2022, \mn@doi [arXiv
  e-prints] {10.48550/arXiv.2212.11991}, \href
  {https://ui.adsabs.harvard.edu/abs/2022arXiv221211991F} {p. arXiv:2212.11991}

\bibitem[\protect\citeauthoryear{{Fuchs}}{{Fuchs}}{2001}]{Fuchs2001}
{Fuchs} B.,  2001, \mn@doi [\aap] {10.1051/0004-6361:20000562}, \href
  {https://ui.adsabs.harvard.edu/abs/2001A&A...368..107F} {368, 107}

\bibitem[\protect\citeauthoryear{{Fuchs}, {Dettbarn}  \& {Tsuchiya}}{{Fuchs}
  et~al.}{2005}]{fuchs2005}
{Fuchs} B.,  {Dettbarn} C.,   {Tsuchiya} T.,  2005, \mn@doi [\aap]
  {10.1051/0004-6361:20052657}, \href
  {https://ui.adsabs.harvard.edu/abs/2005A&A...444....1F} {444, 1}

\bibitem[\protect\citeauthoryear{{Gaia Collaboration} et~al.,}{{Gaia
  Collaboration} et~al.}{2018a}]{gaiadr2_summary}
{Gaia Collaboration} et~al., 2018a, \mn@doi [\aap]
  {10.1051/0004-6361/201833051}, \href
  {https://ui.adsabs.harvard.edu/abs/2018A\&A...616A...1G} {616, A1}

\bibitem[\protect\citeauthoryear{{Gaia Collaboration} et~al.,}{{Gaia
  Collaboration} et~al.}{2018b}]{gaiadr2_vrad}
{Gaia Collaboration} et~al., 2018b, \mn@doi [\aap]
  {10.1051/0004-6361/201832865}, \href
  {https://ui.adsabs.harvard.edu/abs/2018A\&A...616A..11G} {616, A11}

\bibitem[\protect\citeauthoryear{{Ghosh}, {Debattista}  \&
  {Khachaturyants}}{{Ghosh} et~al.}{2022}]{ghosh2022}
{Ghosh} S.,  {Debattista} V.~P.,   {Khachaturyants} T.,  2022, \mn@doi [\mnras]
  {10.1093/mnras/stac137}, \href
  {https://ui.adsabs.harvard.edu/abs/2022MNRAS.511..784G} {511, 784}

\bibitem[\protect\citeauthoryear{{Goldreich} \& {Lynden-Bell}}{{Goldreich} \&
  {Lynden-Bell}}{1965}]{goldreich1965}
{Goldreich} P.,  {Lynden-Bell} D.,  1965, \mn@doi [\mnras]
  {10.1093/mnras/130.2.97}, \href
  {https://ui.adsabs.harvard.edu/abs/1965MNRAS.130...97G} {130, 97}

\bibitem[\protect\citeauthoryear{{Goldreich} \& {Tremaine}}{{Goldreich} \&
  {Tremaine}}{1978}]{goldreich1978}
{Goldreich} P.,  {Tremaine} S.,  1978, \mn@doi [\apj] {10.1086/156203}, \href
  {https://ui.adsabs.harvard.edu/abs/1978ApJ...222..850G} {222, 850}

\bibitem[\protect\citeauthoryear{{G{\'o}mez} et~al.,}{{G{\'o}mez}
  et~al.}{2012}]{gomez2012}
{G{\'o}mez} F.~A.,  et~al., 2012, \mn@doi [\mnras]
  {10.1111/j.1365-2966.2012.21176.x}, \href
  {https://ui.adsabs.harvard.edu/abs/2012MNRAS.423.3727G} {423, 3727}

\bibitem[\protect\citeauthoryear{Harris et~al.,}{Harris
  et~al.}{2020}]{harris2020}
Harris C.~R.,  et~al., 2020, \mn@doi [Nature] {10.1038/s41586-020-2649-2}, 585,
  357

\bibitem[\protect\citeauthoryear{{Hawley}, {Gammie}  \& {Balbus}}{{Hawley}
  et~al.}{1995}]{hawley1995}
{Hawley} J.~F.,  {Gammie} C.~F.,   {Balbus} S.~A.,  1995, \mn@doi [\apj]
  {10.1086/175311}, \href
  {https://ui.adsabs.harvard.edu/abs/1995ApJ...440..742H} {440, 742}

\bibitem[\protect\citeauthoryear{Hill}{Hill}{1878}]{hill1878}
Hill G.~W.,  1878, American journal of Mathematics, 1, 5

\bibitem[\protect\citeauthoryear{{Hunt}, {Price-Whelan}, {Johnston}  \&
  {Darragh-Ford}}{{Hunt} et~al.}{2022}]{hunt2022}
{Hunt} J. A.~S.,  {Price-Whelan} A.~M.,  {Johnston} K.~V.,   {Darragh-Ford} E.,
   2022, \mn@doi [\mnras] {10.1093/mnrasl/slac082}, \href
  {https://ui.adsabs.harvard.edu/abs/2022MNRAS.516L...7H} {516, L7}

\bibitem[\protect\citeauthoryear{{Ibata}, {Gilmore}  \& {Irwin}}{{Ibata}
  et~al.}{1994}]{ibata1994}
{Ibata} R.~A.,  {Gilmore} G.,   {Irwin} M.~J.,  1994, \mn@doi [\nat]
  {10.1038/370194a0}, \href
  {https://ui.adsabs.harvard.edu/abs/1994Natur.370..194I} {370, 194}

\bibitem[\protect\citeauthoryear{{Johnston}, {Spergel}  \&
  {Hernquist}}{{Johnston} et~al.}{1995}]{johnston1995}
{Johnston} K.~V.,  {Spergel} D.~N.,   {Hernquist} L.,  1995, \mn@doi [\apj]
  {10.1086/176247}, \href
  {https://ui.adsabs.harvard.edu/abs/1995ApJ...451..598J} {451, 598}

\bibitem[\protect\citeauthoryear{{Julian} \& {Toomre}}{{Julian} \&
  {Toomre}}{1966}]{julian1966}
{Julian} W.~H.,  {Toomre} A.,  1966, \mn@doi [\apj] {10.1086/148957}, \href
  {https://ui.adsabs.harvard.edu/abs/1966ApJ...146..810J} {146, 810}

\bibitem[\protect\citeauthoryear{{Kalnajs}}{{Kalnajs}}{1973}]{kalnajs1973}
{Kalnajs} A.~J.,  1973, \mn@doi [\apj] {10.1086/152023}, \href
  {https://ui.adsabs.harvard.edu/abs/1973ApJ...180.1023K} {180, 1023}

\bibitem[\protect\citeauthoryear{{Khoperskov}, {Di Matteo}, {Gerhard}, {Katz},
  {Haywood}, {Combes}, {Berczik}  \& {Gomez}}{{Khoperskov}
  et~al.}{2019}]{khoperskov2019}
{Khoperskov} S.,  {Di Matteo} P.,  {Gerhard} O.,  {Katz} D.,  {Haywood} M.,
  {Combes} F.,  {Berczik} P.,   {Gomez} A.,  2019, \mn@doi [\aap]
  {10.1051/0004-6361/201834707}, \href
  {https://ui.adsabs.harvard.edu/abs/2019A&A...622L...6K} {622, L6}

\bibitem[\protect\citeauthoryear{{King}}{{King}}{1966}]{king1966}
{King} I.~R.,  1966, \mn@doi [\aj] {10.1086/109857}, \href
  {https://ui.adsabs.harvard.edu/abs/1966AJ.....71...64K} {71, 64}

\bibitem[\protect\citeauthoryear{{Kumar}, {Ghosh}, {Kataria}, {Das}  \&
  {Debattista}}{{Kumar} et~al.}{2022}]{kumar2022}
{Kumar} A.,  {Ghosh} S.,  {Kataria} S.~K.,  {Das} M.,   {Debattista} V.~P.,
  2022, \mn@doi [\mnras] {10.1093/mnras/stac2302}, \href
  {https://ui.adsabs.harvard.edu/abs/2022MNRAS.516.1114K} {516, 1114}

\bibitem[\protect\citeauthoryear{{Laporte}, {Minchev}, {Johnston}  \&
  {G{\'o}mez}}{{Laporte} et~al.}{2019}]{laporte2019}
{Laporte} C. F.~P.,  {Minchev} I.,  {Johnston} K.~V.,   {G{\'o}mez} F.~A.,
  2019, \mn@doi [\mnras] {10.1093/mnras/stz583}, \href
  {https://ui.adsabs.harvard.edu/abs/2019MNRAS.485.3134L} {485, 3134}

\bibitem[\protect\citeauthoryear{{Li} \& {Shen}}{{Li} \& {Shen}}{2020}]{Li2020}
{Li} Z.-Y.,  {Shen} J.,  2020, \mn@doi [\apj] {10.3847/1538-4357/ab6b21}, \href
  {https://ui.adsabs.harvard.edu/abs/2020ApJ...890...85L} {890, 85}

\bibitem[\protect\citeauthoryear{{Li} \& {Widrow}}{{Li} \&
  {Widrow}}{2021}]{LW2021}
{Li} H.,  {Widrow} L.~M.,  2021, \mn@doi [\mnras] {10.1093/mnras/stab574},
  \href {https://ui.adsabs.harvard.edu/abs/2021MNRAS.503.1586L} {503, 1586}

\bibitem[\protect\citeauthoryear{{Li} \& {Widrow}}{{Li} \&
  {Widrow}}{2023}]{LW2023}
{Li} H.,  {Widrow} L.~M.,  2023, \mn@doi [\mnras] {10.1093/mnras/stad244},
  \href {https://ui.adsabs.harvard.edu/abs/2023MNRAS.tmp..264L} {}

\bibitem[\protect\citeauthoryear{{Mathur}}{{Mathur}}{1990}]{mathur1990}
{Mathur} S.~D.,  1990, \mnras, \href
  {http://adsabs.harvard.edu/abs/1990MNRAS.243..529M} {243, 529}

\bibitem[\protect\citeauthoryear{{Monari}, {Famaey}  \& {Siebert}}{{Monari}
  et~al.}{2015}]{monari2015}
{Monari} G.,  {Famaey} B.,   {Siebert} A.,  2015, \mn@doi [\mnras]
  {10.1093/mnras/stv1206}, \href
  {https://ui.adsabs.harvard.edu/abs/2015MNRAS.452..747M} {452, 747}

\bibitem[\protect\citeauthoryear{{Monari}, {Famaey}  \& {Siebert}}{{Monari}
  et~al.}{2016}]{monari2016}
{Monari} G.,  {Famaey} B.,   {Siebert} A.,  2016, \mn@doi [\mnras]
  {10.1093/mnras/stw171}, \href
  {https://ui.adsabs.harvard.edu/abs/2016MNRAS.457.2569M} {457, 2569}

\bibitem[\protect\citeauthoryear{{Purcell}, {Bullock}, {Tollerud}, {Rocha}  \&
  {Chakrabarti}}{{Purcell} et~al.}{2011}]{purcell2011}
{Purcell} C.~W.,  {Bullock} J.~S.,  {Tollerud} E.~J.,  {Rocha} M.,
  {Chakrabarti} S.,  2011, \mn@doi [\nat] {10.1038/nature10417}, \href
  {https://ui.adsabs.harvard.edu/abs/2011Natur.477..301P} {477, 301}

\bibitem[\protect\citeauthoryear{{Sch{\"o}nrich} \& {Dehnen}}{{Sch{\"o}nrich}
  \& {Dehnen}}{2018}]{schonrich2018}
{Sch{\"o}nrich} R.,  {Dehnen} W.,  2018, \mn@doi [\mnras]
  {10.1093/mnras/sty1256}, \href
  {https://ui.adsabs.harvard.edu/abs/2018MNRAS.478.3809S} {478, 3809}

\bibitem[\protect\citeauthoryear{{Spitzer}}{{Spitzer}}{1942}]{spitzer1942}
{Spitzer} L.,  1942, \apj, 95

\bibitem[\protect\citeauthoryear{{Toomre}}{{Toomre}}{1964}]{toomre1964}
{Toomre} A.,  1964, \mn@doi [\apj] {10.1086/147861}, \href
  {https://ui.adsabs.harvard.edu/abs/1964ApJ...139.1217T} {139, 1217}

\bibitem[\protect\citeauthoryear{{Toomre}}{{Toomre}}{1981}]{toomre1981}
{Toomre} A.,  1981, in {Fall} S.~M.,  {Lynden-Bell} D.,  eds, Structure and
  Evolution of Normal Galaxies. pp 111--136

\bibitem[\protect\citeauthoryear{{Tremaine}}{{Tremaine}}{1999}]{tremaine1999}
{Tremaine} S.,  1999, \mn@doi [\mnras] {10.1046/j.1365-8711.1999.02690.x},
  \href {https://ui.adsabs.harvard.edu/abs/1999MNRAS.307..877T} {307, 877}

\bibitem[\protect\citeauthoryear{{Tremaine}, {Frankel}  \& {Bovy}}{{Tremaine}
  et~al.}{2022}]{tremaine2022}
{Tremaine} S.,  {Frankel} N.,   {Bovy} J.,  2022, \mn@doi [arXiv e-prints]
  {10.48550/arXiv.2212.11990}, \href
  {https://ui.adsabs.harvard.edu/abs/2022arXiv221211990T} {p. arXiv:2212.11990}

\bibitem[\protect\citeauthoryear{Virtanen et~al.,}{Virtanen
  et~al.}{2020}]{virtanen2020}
Virtanen P.,  et~al., 2020, \mn@doi [Nature Methods]
  {10.1038/s41592-019-0686-2}, \href {https://rdcu.be/b08Wh} {17, 261}

\bibitem[\protect\citeauthoryear{{Weinberg}}{{Weinberg}}{1991}]{weinberg1991}
{Weinberg} M.~D.,  1991, \mn@doi [\apj] {10.1086/170059}, \href
  {https://ui.adsabs.harvard.edu/abs/1991ApJ...373..391W} {373, 391}

\bibitem[\protect\citeauthoryear{{Widmark}}{{Widmark}}{2019}]{Widmark2019}
{Widmark} A.,  2019, \mn@doi [\aap] {10.1051/0004-6361/201834718}, \href
  {https://ui.adsabs.harvard.edu/abs/2019A\&A...623A..30W} {623, A30}

\bibitem[\protect\citeauthoryear{{Widmark}, {Laporte}  \& {de Salas}}{{Widmark}
  et~al.}{2021a}]{widmark2021a}
{Widmark} A.,  {Laporte} C.,   {de Salas} P.~F.,  2021a, \mn@doi [\aap]
  {10.1051/0004-6361/202140650}, \href
  {https://ui.adsabs.harvard.edu/abs/2021A&A...650A.124W} {650, A124}

\bibitem[\protect\citeauthoryear{{Widmark}, {Laporte}, {de Salas}  \&
  {Monari}}{{Widmark} et~al.}{2021b}]{widmark2021b}
{Widmark} A.,  {Laporte} C.~F.~P.,  {de Salas} P.~F.,   {Monari} G.,  2021b,
  \mn@doi [\aap] {10.1051/0004-6361/202141466}, \href
  {https://ui.adsabs.harvard.edu/abs/2021A&A...653A..86W} {653, A86}

\bibitem[\protect\citeauthoryear{{Widmark}, {Widrow}  \& {Naik}}{{Widmark}
  et~al.}{2022}]{widmark2022}
{Widmark} A.,  {Widrow} L.~M.,   {Naik} A.,  2022, \mn@doi [\aap]
  {10.1051/0004-6361/202244453}, \href
  {https://ui.adsabs.harvard.edu/abs/2022A&A...668A..95W} {668, A95}

\bibitem[\protect\citeauthoryear{{Widrow} \& {Bonner}}{{Widrow} \&
  {Bonner}}{2015}]{widrow2015}
{Widrow} L.~M.,  {Bonner} G.,  2015, \mn@doi [\mnras] {10.1093/mnras/stv574},
  \href {https://ui.adsabs.harvard.edu/abs/2015MNRAS.450..266W} {450, 266}

\bibitem[\protect\citeauthoryear{{Widrow}, {Gardner}, {Yanny}, {Dodelson}  \&
  {Chen}}{{Widrow} et~al.}{2012}]{widrow2012}
{Widrow} L.~M.,  {Gardner} S.,  {Yanny} B.,  {Dodelson} S.,   {Chen} H.-Y.,
  2012, \mn@doi [\apjl] {10.1088/2041-8205/750/2/L41}, \href
  {http://adsabs.harvard.edu/abs/2012ApJ...750L..41W} {750, L41}

\bibitem[\protect\citeauthoryear{{Williams} et~al.,}{{Williams}
  et~al.}{2013}]{williams2013}
{Williams} M.~E.~K.,  et~al., 2013, \mn@doi [\mnras] {10.1093/mnras/stt1522},
  \href {http://adsabs.harvard.edu/abs/2013MNRAS.436..101W} {436, 101}

\bibitem[\protect\citeauthoryear{{Xu}, {Newberg}, {Carlin}, {Liu}, {Deng},
  {Li}, {Sch{\"o}nrich}  \& {Yanny}}{{Xu} et~al.}{2015}]{xu2015}
{Xu} Y.,  {Newberg} H.~J.,  {Carlin} J.~L.,  {Liu} C.,  {Deng} L.,  {Li} J.,
  {Sch{\"o}nrich} R.,   {Yanny} B.,  2015, \mn@doi [\apj]
  {10.1088/0004-637X/801/2/105}, \href
  {https://ui.adsabs.harvard.edu/abs/2015ApJ...801..105X} {801, 105}

\bibitem[\protect\citeauthoryear{{Yanny} \& {Gardner}}{{Yanny} \&
  {Gardner}}{2013}]{yanny2013}
{Yanny} B.,  {Gardner} S.,  2013, \mn@doi [\apj] {10.1088/0004-637X/777/2/91},
  \href {https://ui.adsabs.harvard.edu/abs/2013ApJ...777...91Y} {777, 91}

\makeatother
\end{thebibliography}

\end{document}